\documentclass[%
 reprint,
 amsmath,amssymb,
 aps, pra
]{revtex4-2}

\usepackage{graphicx} 
\usepackage{dcolumn}
\usepackage{bm} % bold math
\usepackage{xcolor} % Colors
\usepackage{float} % Figures
\usepackage{subcaption}
\usepackage{tikz} % TikZ
\usetikzlibrary{quantikz} % Quantum circuits
\usepackage{standalone}
\usepackage{pgfplots}
\usepackage{caption} 
\usepackage{hyperref}

\pgfplotsset{compat=1.18}

\begin{document}

\preprint{APS/123-QED}

% \title{Pauli Check Extrapolation for Quantum Error Mitigation}

% \title{Scalable Quantum Error Mitigation through Extrapolation}

% \title{Pauli Check Extrapolation for Expectation Value Estimation}

% \title{Extrapolating Expectation Values through Pauli Checks}

\title{Extrapolating Pauli Checks for Expectation Value Estimation on \\ Noisy Quantum Devices}

\author{Quinn Langfitt$^{1,3}$}
\thanks{These authors contributed equally.}
\author{Ji Liu$^1$}
\thanks{These authors contributed equally.}
\author{Benchen Huang$^2$}
\author{Alvin Gonzales$^1$}
\author{Kaitlin N. Smith$^3$}
\author{Nikos Hardavellas$^{3,4}$}
\author{Zain H. Saleem$^1$}

\affiliation{$^1$Mathematics and Computer Science Division, 
Argonne National Laboratory, Lemont, IL, USA}

\affiliation{$^2$Department of Chemistry, University of Chicago, Chicago, IL, USA}

\affiliation{$^3$Department of Computer Science, Northwestern University, Evanston, IL, USA}

\affiliation{$^4$Department of Electrical and Computer Engineering, Northwestern University, Evanston, IL, USA}

\begin{abstract}
    Pauli Check Sandwiching (PCS) is an error detection scheme that protects quantum circuits by inserting pairs of parity checks and discarding runs that signal errors. However, each additional check introduces noise and exponentially increases sampling costs. To address these limitations, we propose Pauli Check Extrapolation (PCE), an error mitigation technique that obtains measured expectation values from circuits with different numbers of checks and, analogous to ZNE, extrapolates to the “maximum check” limit — the theoretical number of checks required for unit fidelity. We test linear and exponential ansatzes, deriving the exponential form from the Markovian error model. Benchmarking PCE against ZNE on random Clifford circuits with simulated depolarizing noise shows PCE outperforming ZNE for larger circuits. On real IBM hardware, PCE achieves an accuracy of up to $99.2\%$ ($56.2\%$ improvement over baseline), compared to ZNE's $82\%$ accuracy ($29.1\%$ improvement over baseline), for $4$-qubit circuits. To demonstrate a practical use case, we then apply PCE towards mitigating errors in classical shadow measurements. Our results show that PCE can achieve fidelities greater than the state-of-the-art Robust Shadow estimation, while significantly reducing the number of required samples by eliminating the need for a calibration procedure. We validate these findings on both fully connected topologies and simulated IBM hardware backends.
\end{abstract}

\maketitle

%\tableofcontents

\section{Introduction}

Recent experimental breakthroughs have demonstrated quantum computers achieving computational advantage over classical devices on carefully designed tasks, often with the help of error mitigation techniques \cite{aharonov2025importanceerrormitigationquantum, zhang2025demonstratingquantumerrormitigation, He_2025, Liao_2024}. These recent demonstrations highlight the rapid progress of quantum technologies and bring practical quantum advantage closer to reality. However, modern-day devices remain limited by decoherence and hardware imperfections, making quantum error mitigation essential for reliable computation in practical applications spanning machine learning \cite{Schuld2015}, combinatorial optimization \cite{farhi2014quantum}, quantum simulation \cite{Li2017}, and cryptography \cite{Buhrman_2014}. Various error mitigation methods have been developed to address these limitations, including zero-noise extrapolation (ZNE) \cite{Temme2017, Kim_2023, Kandala_2019}, probabilistic error cancellation \cite{Endo_2018}, Pauli Check Sandwiching (PCS) \cite{Debroy_2020, Gonzales_2023}, symmetry verification \cite{Bonet_Monroig_2018, McArdle_2019, Shaydulin_2021, Cai_2021}, and simulated quantum error mitigation \cite{liu2022classical}.

Among the aforementioned error mitigation approaches, PCS is unique in that it is an error detection protocol that uses post-selection to provide an error-mitigated output. It works by imposing constraints on the circuit and post-selecting based on outputs that violate those constraints. To implement these constraints, Pauli gates are applied at both ends of the circuit with an ancilla qubit appended for each pair of checks. Since the checks are applied in a layering fashion on the original circuit, each pair of checks is typically referred to as a `layer.' Each sample is then followed by a measurement of the ancilla qubit(s), where only the results corresponding to zero(s) on the ancilla(s) are kept, therefore acting as a post-selection scheme.  In the limit that the checks are ideal, Proposition 2 in \cite{Gonzales_2023} states that there exist $2n$ number of PCS checks (possibly employing checks not in the Pauli group), where $n$ is the number of compute qubits, such that the postselected state is noiseless. The checks consist of left and right pairs and we can always employ weight-one Pauli checks for one of the checks for each pair. 

On the other hand, ZNE is a more traditional error mitigation approach that is designed to mitigate the error of the measured expectation value of an observable. ZNE reduces error by artificially amplifying circuit noise through the insertion of logical identity operations, and then extrapolating the measured expectation values back to the zero-noise limit. 

In this work, we introduce Pauli Check Extrapolation (PCE), an error mitigation technique that extrapolates PCS expectation values to the `maximum check' limit. Here we test linear and exponential models for extrapolation, performing least-squares fitting to the expectation values obtained from different numbers of physically implemented checks. Like ZNE, PCE mitigates errors in measured observables through extrapolation. However, a crucial distinction is that PCE reduces noise through post-selection, whereas ZNE requires artificially increasing noise levels—an approach that fails for already highly noisy circuits. Furthermore, extrapolation methods are typically more accurate when the collected data points are near the target domain value (zero noise or max checks). Thus, a crucial advantage of PCE over ZNE is that it does not require sampling circuits with a significantly larger gate count than the original circuit that is being mitigated. Our approach also avoids the noise typically associated with the physical implementation of additional check layers and eliminates the exponentially increasing sampling overhead with each added layer. Additionally, for non-Clifford circuits where the number of applicable Pauli checks is limited, we can extrapolate to checks that cannot be physically performed, thus bypassing this limitation.

We evaluate PCE through two complementary sets of experiments. First, we benchmark PCE against ZNE on random Clifford circuits with varying qubit counts and circuit depths. Our results demonstrate that while ZNE performs better for smaller circuit sizes, PCE substantially outperforms ZNE for larger circuits (achieving up to $\simeq0.13$ error magnitude improvement). Moreover, PCE maintains consistent extrapolation parameters across all circuit configurations, whereas ZNE requires circuit-specific parameter tuning (e.g.,, extrapolation model, scale factors, folding technique, etc.)---a significant practical advantage when the optimal parameters cannot be determined a priori. On real IBM hardware, PCE achieves a substantial improvement over ZNE, reaching an accuracy of up to $99.2\%$ compared to ZNE's $82\%$.

We then apply PCE to improve the accuracy of classical shadow estimation for variational quantum eigensolver (VQE) circuits~\cite{peruzzo2014variational}, comparing our results with the state-of-the-art Robust Shadow (RS) estimation \cite{Chen_2021}. Our simulations contain depolarizing noise affecting each single-qubit and two-qubit gate, including the checks. Under fully connected qubit topologies, we find that PCE significantly improves fidelity over RS estimation for a 4-qubit H$_2$ VQE circuit when the entire circuit (including state preparation and global Clifford portions) is protected. We also demonstrate improved fidelities for the $8$-qubit H$_2$O VQE circuit under homogeneous and inhomogeneous noise distributions. Furthermore, for our experiments PCE showed comparable, and sometimes improved, performance on fake IBM hardware backends, depending on the device connectivity.

 Another advantage of PCE over RS estimation is that it does not require a calibration procedure. By eliminating this calibration step, PCE offers the additional benefit of reducing the total required sampling size. This advantage is significant, even if the fidelity is slightly lower for some instances.

\section{Background}
\label{Sec:Background}
\subsection{Pauli Check Sandwiching}

PCS sandwiches the target computational circuit, denoted by \(U\), between pairs of controlled Pauli gates. Mathematically, these Pauli gates are chosen from the $\pm 1$ elements of the Pauli group \(\mathcal{P}_n = \{I, X, Y, Z\}^{\otimes n} \times \{ \pm 1, \pm i \}\). The controlled Pauli gates, \(L, R \in \mathcal{P}_n\), are selected such that

\begin{equation}
\label{eq:PCS}
    L U R = U.
\end{equation}

\begin{figure}[h!]
\centering
\includegraphics[width=0.45\textwidth]{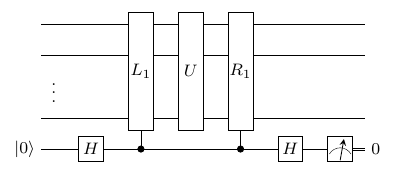}
\caption{Pauli Check Sandwiching circuit with a single layer of checks. Here $U$ is the compute circuit.} 
\label{fig:PCS_single}
\end{figure}

To detect errors, an ancilla qubit is appended to the system for each layer of Pauli checks. The detection of an error is indicated by a measurement outcome of \(1\) on the ancilla qubit(s). Conversely, a measurement outcome of \(0\) on all ancilla qubits signifies no detected errors, and thus, the results of the computation are kept. Through this process of post-selection, PCS effectively mitigates the impact of errors in the computational circuit. An example of a circuit with one layer of Pauli checks is shown in Fig. ~\ref{fig:PCS_single}.

\begin{figure}[h!]
\centering
\includegraphics[width=0.48\textwidth]{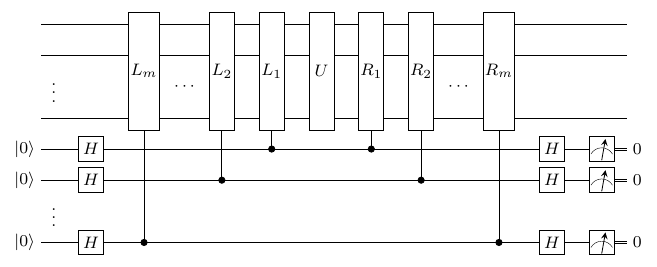}
\caption{
Pauli Check Sandwiching circuit with $m$ layers applied to the payload circuit $U$. 
}
\label{fig:PCS_multi}
\end{figure}

We can introduce additional layers to our circuit, where each pair of Pauli gates $L_{n}$, $R_{n}$, where $n$ is the $n$th layer, satisfies the condition in Eq. ~\ref{eq:PCS}. This will theoretically allow the circuit to detect more errors and therefore improve the state fidelity. Furthermore, as stated in Proposition $2$ of \cite{Gonzales_2023}, in the theoretical limit of noiseless checks, there exists a set of checks (which may include non-Pauli checks for general circuits) that ensures the post-selected state is noiseless. An example of a circuit with multiple layers is shown in Fig. ~\ref{fig:PCS_multi}.

\subsection{Classical Shadow Tomography}

The goal of classical shadow tomography~\cite{Huang_2020} is to predict the expectation value of $M$ physical observables for a certain quantum state (simultaneously) from relatively few measurements. It works by i). performing random measurements over the quantum state and ii). estimating all the expectation values through post-processing measurement outcomes. Shadow tomography is an approximate theory, and it requires a number of samples:
\[N = O\left(\log(M) \max_i \left\|O_i\right\|^2_{\text{shadow}} / \epsilon^2\right),\]
to keep errors for the estimation of each observable within $\epsilon$. In the above equation, \(\left\|O_i\right\|_{\text{shadow}}\) denotes the shadow norm of the \(i\)-th observable, which depends on the set of randomized measurements we use. What's essential about this protocol is that the number of samples required only scales logarithmically with $M$, making it promising for practical quantum simulation applications~\cite{huggins2022unbiasing, o2022efficient, avdic2023fewer}. For our experiments, we use the global Clifford group $\text{Cl}(2^n)$, which has a shadow norm of \(\left\|O_i\right\|_{\text{shadow}} < 3 \operatorname{tr}(O_i^2)\).
%, whereas sampling from a set of tensor products of local unitaries has a shadow norm of \(\left\|O_i \right\|_{\text{shadow}} \leq 4^w \left\|O_i\right\|_{\infty}\).

To apply PCS to classical shadow tomography, we can sandwich the global Clifford at the end of the circuit. For any given Pauli check \(P_2\), there will always exist a corresponding Pauli check \(P_1\) for a circuit with only Clifford gates. This configuration is theoretically guaranteed to protect that portion of the circuit up to unit fidelity if the checks are noiseless (Theorem 1 in \cite{Gonzales_2023}). Alternatively, we can include both the state preparation circuit and the random global Clifford. An advantage of extrapolation is that, when including the state preparation circuit—potentially a non-Clifford circuit—we can extrapolate to checks that cannot be physically applied. This is important because the number of pairs \(L\) and \(R\) that satisfy Eq.~\ref{eq:PCS} decreases with the number of non-Clifford gates. Moreover, for larger quantum circuits, we can theoretically guarantee finding pairs of Pauli checks for portions of the circuit by focusing on protecting blocks of Clifford circuits. We present results for both configurations in Section \ref{sec:Results}.

% In both cases, we observe that the PCS scheme performs as well, if not better, than the state-of-the-art robust shadow estimation scheme. Furthermore, PCS performs better when it includes both the state preparation and the Clifford portions of the circuit than just the Clifford portion alone. However, one challenge in including the entire circuit is that it is not guaranteed that we will find Pauli checks for an arbitrary circuit. Therefor e, it can take longer to search through the space of Pauli checks that satisfies the constraint \(P_1UP_2 = U\).

\subsection{Robust Shadow Estimation}

RS estimation \cite{Chen_2021, Koh_2022} is a protocol designed for mitigating errors in the shadow circuit, originally for global/local Clifford group $\text{Cl}(2^n), \text{Cl}_2^{\otimes n}$, and recently generalized to the fermionic case~\cite{zhao2023group, wu2024error}. This protocol could in principle recover the noise-free expectation value if i). the state preparation is perfect, and ii). the shadow circuit noise is gate-independent, time-stationary, and Markovian (GTM). %This scheme works by first characterizing the noisy channel of classical shadow that ultimately compensates for the effect of noise during post-processing.

RS estimation operates by first characterizing the noisy channel of classical shadow estimation, which is ultimately used to compensate for the effect of noise during post-processing. This characterization is accomplished through a calibration procedure involving sampling over shadow circuits. This calibration contributes to a significant sampling overhead compared to PCS, which does not require such a step. Theorem $7$ in Ref.~\cite{Chen_2021} states that when sampling from the global Clifford group, the number of total required samples is
\[
R = 136 \ln(2\delta^{-1}) \frac{(1 + \varepsilon^2) (1 + \frac{1}{d})^2}{\varepsilon^2 (F_Z - \frac{1}{d})^2},
\]
where $\varepsilon$ sets the threshold for calibrating the noise channel with a success probability of at least $1-\delta$, $d$ represents the dimension of the problem, and $F_Z$ is called the $Z$-basis fidelity of the noise channel which can be approximated as unity when the noise is small~\cite{Chen_2021}. Although we note that the sampling complexity is (approximately) independent of system size when the noise is small, in practice this calibration step can introduce a fairly large sampling overhead. For example, if we set $\varepsilon = 1\%$, this increases the number of samples by a factor of $1/\varepsilon^2 = 10,000$ to the already existing $136$ pre-factor. These additional calibration samples can be challenging to implement on real hardware.

\section{Methodology}

\subsection{Pauli Check Extrapolation}

PCE is intended to address three core challenges associated with the implementation of PCS:

\begin{enumerate}
    \item On real hardware, each additional check introduces more noise to the circuit.
    \item The post-selection success rate decreases exponentially with the number of layers.
    \item For non-Clifford circuits, the number of applicable Pauli checks is limited.
\end{enumerate}

To address these limitations, we propose an extrapolation approach that simulates the effect of maximizing the number of PCS layers without physically implementing them. This maximum layer count is tailored to the system size and the specific observables required; for instance, only incorporating a controlled Pauli Z gate for each qubit when measurements are solely in the Z-basis. Extrapolation effectively bypasses the additional noise and sampling overhead that would result from actually implementing additional layers. Additionally, for non-Clifford circuits where the number of applicable Pauli checks is limited, extrapolation allows us to bypass this limitation, provided that we can at least find a few checks.

To perform PCE, the first $m$ layers of Pauli checks are implemented on the quantum circuit and the expectation value of the target observable is measured for each configuration. This yields a dataset $\{(n, E_n)\}_{n=1}^{m}$, where $n$ represents the number of check layers and $E_n$ is the corresponding expectation value. 

An extrapolation model $E(n)$ is then fitted to the collected data. Two primary models are considered in this work:
\begin{itemize}
    \item Linear: $E(n) = \alpha + \beta n$
    \item Exponential: $E(n) = ab^n + c$
\end{itemize}
where $\alpha, \beta$ (linear) and $a, b, c$ (exponential) are fitting parameters determined by minimizing the sum of squared residuals:
\begin{equation}
    \min_{\theta} \sum_{n=1}^{m} (E_n - E(n; \theta))^2
\end{equation}
where $\theta$ represents the model parameters.

Finally, the expectation value is extrapolated to $n_{\text{max}}$, the theoretical number of checks required for unit fidelity. For an $n$-qubit system with measurements in the computational basis, $n_{\text{max}} = n$ when using single-qubit $Z$ checks. More generally, $n_{\text{max}}$ depends on the measurement basis and can extend up to $2n$ single-weight checks for arbitrary observables \cite{Gonzales_2023}. The extrapolated value $E(n_{\text{max}})$ provides the error-mitigated estimate of the observable expectation value. Figure~\ref{fig:extrapolation_ex} illustrates this extrapolation process.

\begin{figure}[htbp]
    \centering
    \hspace*{-1cm} % Adjust this value to control the offset
    \begin{tikzpicture}
    % Set background color
    % \fill[lightgray!30] (0,0) rectangle (10,6);

    \begin{axis}[
        xlabel={Number of Checks},
        ylabel={Expectation Value},
        grid=major,
        width=10cm,
        height=6cm,
        xmin=1, xmax=4,
        ymin=0, ymax=1.2,
        domain=1:4,
        samples=100,
        legend pos=south east,
        ytick={0, 0.2, 0.4, 0.6, 0.8, 1.0, 1.2},
        xtick={1, 2, 3, 4},
        axis background/.style={fill=cyan!10}, % Light cyan background for axis
        major grid style={line width=.2pt, draw=gray!50},
        minor grid style={line width=.1pt, draw=gray!30},
        axis line style={thick, color=black}, % Thicker axis lines
        tick style={color=black}, % Black ticks for better visibility
        label style={font=\large} % Larger font for labels
    ]
    % Plot the linear fit
    \addplot [gray, thick] {0.2*(x-1) + 0.4};
    \addlegendentry{Linear Fit}
    
    % Plot the first three data points in blue, slightly offset, with circles
    \addplot [
        only marks,
        mark=*,
        mark options={fill=blue},
        blue,
        mark size=2
    ] coordinates {
        (1, 0.45)
        (2, 0.65)
        (3, 0.75)
    };
    \addlegendentry{Measured Checks}
    
    % Plot the last data point in blue, with a big diamond marker
    \addplot [
        only marks,
        mark=diamond*,
        mark options={fill=red},
        red,
        mark size=4
    ] coordinates {
        (4, 1.0)
    };
    \addlegendentry{Extrapolated Check}
    
    \end{axis}
\end{tikzpicture}
    \caption{Schematic of a linear model fitted to the expectation values of the physically implemented checks (blue circle) and extrapolating to the fourth check (red diamond).}
    \label{fig:extrapolation_ex}
\end{figure}

\subsection{Expectation Value Ansatz}
\label{sec:exp_ansatz}
We can use the Markovian error model described in Ref.~\cite{Vandenberg_2023SingleShotErrMitigByCohPauliChecks} to derive an ansatz for the expectation value. Let $\pi_1$, $\pi_2$, and $\pi_3$ denote the states of detected error, undetected error, and no error, respectively. Let $\epsilon$ denote the probability of an error on the data qubits. Initially (without checks), the state of the quantum computer is described by the vector
\begin{align}
    \vec{\pi}=\begin{pmatrix}
        \pi^{(0)}_1\\
        \pi^{(0)}_2\\
        \pi^{(0)}_3
    \end{pmatrix}=\begin{pmatrix}
        0\\
        \epsilon\\
        1-\epsilon
    \end{pmatrix}
\end{align}
and the transition matrix for each check is given by the upper triangular matrix
\begin{align}
    T=\begin{pmatrix}
    1 &\frac{1}{2} &t_d\\
    0 &\frac{1}{2} &t_u\\
    0 & 0 & t_{ok}
    \end{pmatrix},
\end{align}
where $t_d$, $t_u$, $t_{ok}$ are the probabilities of the check introducing a detectable error, an undetectable error, and no error, respectively \cite{Vandenberg_2023SingleShotErrMitigByCohPauliChecks}. The probability of a logical error is given by
\begin{align}
    P(\text{logical error})=\frac{\pi_2}{\pi_2+\pi_3}.
\end{align}
Since we are interested in the logical error, we can ignore the detected error state and focus on the block matrix
\begin{align}
    T'=\begin{pmatrix}
    \frac{1}{2} &t_u\\
    0 & t_{ok}
    \end{pmatrix}.
\end{align}
% Let $t_u\approx 0$. This implies that $p\approx 0$, $t_d\approx 0$, and $t_{ok}\approx 1$. The 
% The transition matrix for $m$ checks is
% \begin{align}
%     T'^k=\begin{pmatrix}
%     \left(\frac{1}{2}\right)^m &\left(\frac{1}{2}\right)^{m-1}t_u\\
%     0 & t_{ok}^m
%     \end{pmatrix}+\begin{pmatrix}
%     0 &O(t_ut_{ok})\\
%     0 & 0
%     \end{pmatrix}
% \end{align}
% The logical error rate for $k$ checks is given by
% \begin{align}
%     P(\text{logical error})=\frac{\left(\frac{1}{2}\right)^m\epsilon + \left(\frac{1}{2}\right)^{m-1}t_u(1-\epsilon)}{t_{ok}^m(1-\epsilon)+\left(\frac{1}{2}\right)^m\epsilon + \left(\frac{1}{2}\right)^{m-1}t_u(1-\epsilon)}
% \end{align}
Let the checks be near perfect, i.e., $t_d= 0$, $t_u= 0$, and $t_{ok}= 1$. The logical error rate after $m$ checks for this scenario is
\begin{align}
    P(\text{logical error})=\frac{\left(\frac{1}{2}\right)^m\epsilon}{1-\epsilon+\left(\frac{1}{2}\right)^m\epsilon}=\frac{\epsilon}{2^m(1-\epsilon)+\epsilon},
\end{align}
which is exponential in $m$. Therefore, the expectation value as a function of the number of checks $m$ should also be exponential.  
%and we can use the exponential ansatz $E(m)=a\left(\frac{1}{2}\right)^m+c$, where $a$ and $c$ are scalars determined from fitting. 
In this paper, we use the form
\begin{align}
\label{eq:exp_ansatz}
    E(m)=ab^m+c,
\end{align}
where $a$, $b$, and $c$ are scalars determined from fitting. This result is intuitive because the set of undetectable commuting Pauli strings exponentially decreases (by a factor of $\frac{1}{2})$ with each check.

\section{Results}

\subsection{Benchmarks on Clifford Circuits (PCE vs ZNE)}
\label{sec:ZNE_Comparison}

% \subsection{Experimental Setup}

\textbf{Experimental setup.} In this section, we benchmark the performance of PCE against ZNE on random Clifford circuits. We evaluated two experimental settings: i) simulations with a depolarizing noise model, and ii) executions on IBM quantum hardware. The random Clifford circuits were generated by specifying a circuit depth, then, at each depth, randomly selecting and applying one- or two-qubit Clifford gates to random qubit subsets. For the first set of experiments, the single- and two-qubit gate error rates were set to $5 \times 10^{-4}$ and $5 \times 10^{-3}$, respectively. For the hardware experiments, we used mirrored versions of the random Clifford circuits and executed them on the IBM Kingston backend. We exclusively used Pauli Z checks for every circuit in both sets of experiments. We used the software package mitiq \cite{mitiq} to implement ZNE.

For both sets of experiments, performance was evaluated using the absolute error from the ideal expectation value of the $Z^{\otimes n}$ observable. We restricted attention to circuits with ideal values of $\pm 1$, since noise channels typically drive the measured observable toward zero. In the first set, we randomly sampled Clifford circuits until obtaining 20 instances with $\langle Z^{\otimes n}\rangle = 1$ for each qubit count and circuit depth. The mirrored circuits, which are equivalent to the identity, automatically yield $\langle Z^{\otimes n}\rangle = 1$. For ZNE, we applied global folding in both sets of experiments.

% \subsection{Simulation Results}

\textbf{Simulation results.} For the simulation runs, we extrapolate expectation values using a number of checks equal to half the system size (e.g., 6 checks for 12-qubit circuits). For ZNE experiments, we performed an extensive parameter scan including Richardson, linear, and exponential extrapolation models with the following scale factor sets: [1, 1.1, 1.2], [1, 1.2, 1.6], [1, 3, 5], [1, 2, 3, 4, 5], [1, 3, 5, 7, 9], [1, 1.1, 1.2, 1.3, 1.4], and [1, 1.2, 1.5, 1.8, 2]. The total shot budget was fixed at 50,000 shots, distributed evenly among intermediate circuits. For PCE, shots were divided equally among checked circuits, while for ZNE, they were divided equally among noise-scaled circuits. We averaged the expectation values over $20$ random circuits. 

\begin{figure}[ht]
\centering
\includegraphics[width=0.48\textwidth]{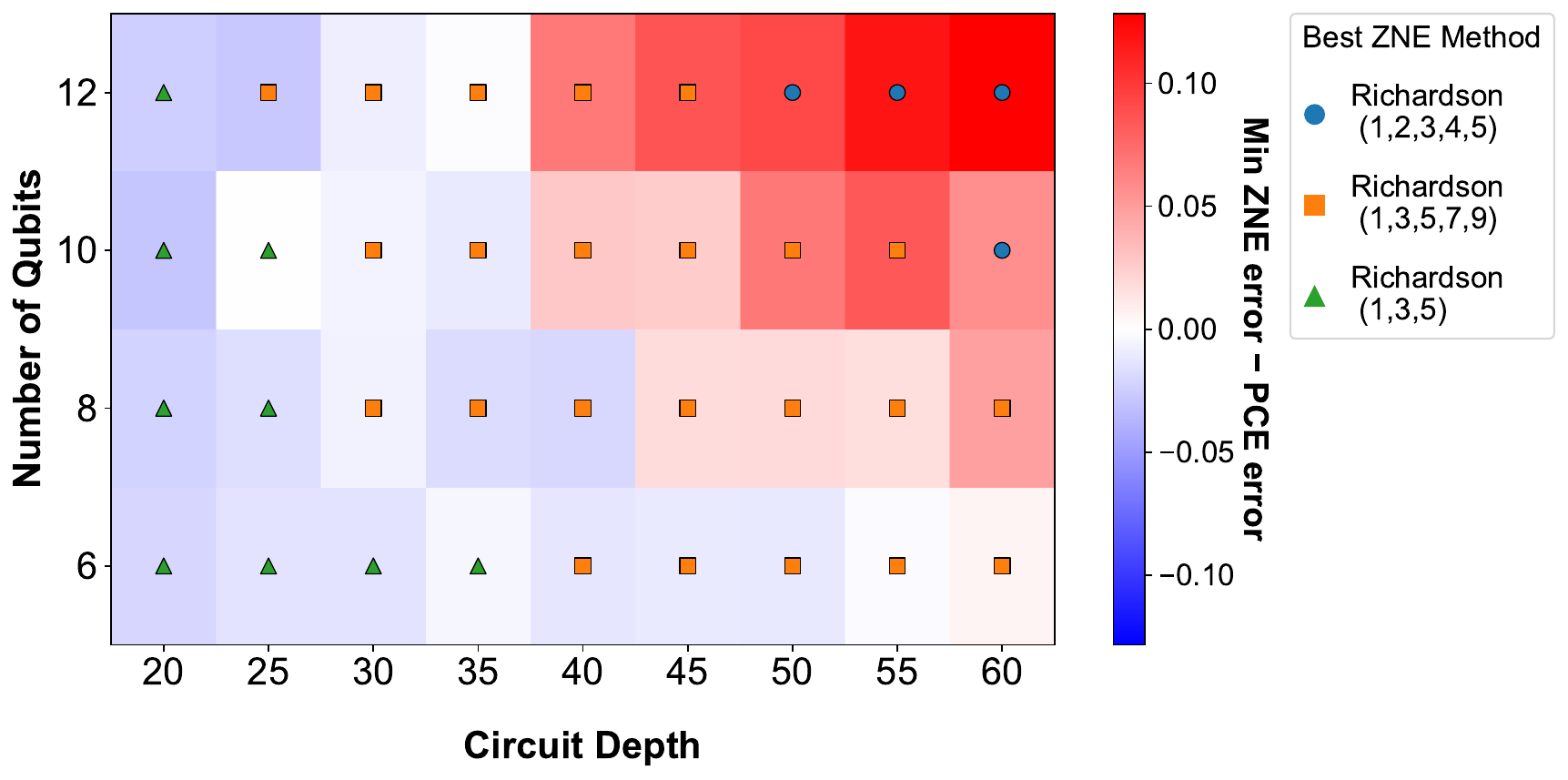}
\caption{Classical simulation. Absolute error difference between the best ZNE method and PCE error using the exponential model across varying circuit depths and qubit counts. Colored shapes indicate the best ZNE method (e.g., Richardson with scale factors (1, 2, 3, 4, 5)) for a given qubit count and circuit depth.}
\label{fig:zne_vs_pce_heatmap}
\end{figure}

The results from the simulated noise backend are summarized in Fig.~\ref{fig:zne_vs_pce_heatmap}. The figure shows the absolute error difference between the best ZNE method tested (i.e., scale factors, extrapolation model) with minimum error and the PCE error using the exponential model. A clear trend emerges: for smaller circuit sizes (both in terms of qubit count and circuit depth), ZNE performs better on average than PCE. However, for larger circuits, PCE outperforms ZNE substantially (up to $\simeq 0.13$ error magnitude difference). Furthermore, a known issue with ZNE is the difficulty of determining in advance which ZNE parameters will perform best. We also observe this in Fig.~\ref{fig:zne_vs_pce_heatmap}, where the best-performing ZNE method, varies across circuit sizes. In contrast, the PCE exponential model we used remains consistent across all circuits.

% \subsection{Hardware Results}

\textbf{Hardware results.} For the hardware experiments, we compare PCE to ZNE using scale factors of $[1, 3, 5]$ with linear and Richardson extrapolation models. The test circuits consist of $4$-qubit mirrored random Clifford circuits. We evaluate PCE extrapolation performance using $2$ and $3$ checks. All measured expectation values are averaged over $5$ random circuits. Each circuit configuration—baseline, PCE with varying check counts, and ZNE with different scale factors—was executed with $50,000$ shots. 

We used \texttt{mapomatic} \cite{PRXQuantum.4.010327} to map the logical qubits of the payload circuit to physical qubits on the device. This mapped circuit served as the baseline implementation and was also used for the ZNE variants. After mapping the payload, we identified candidate ancilla qubits adjacent to the computation qubits and selected those with the lowest two-qubit error rates relative to the target qubits being checked. For the error-mitigated circuits, we applied additional techniques, including the \textit{mthree} readout error mitigation technique \cite{M3}, Pauli twirling \cite{pauli_twirling}, and TREX \cite{van_den_Berg_2022}.

\begin{figure}[ht]
\centering
\includegraphics[width=0.48\textwidth]{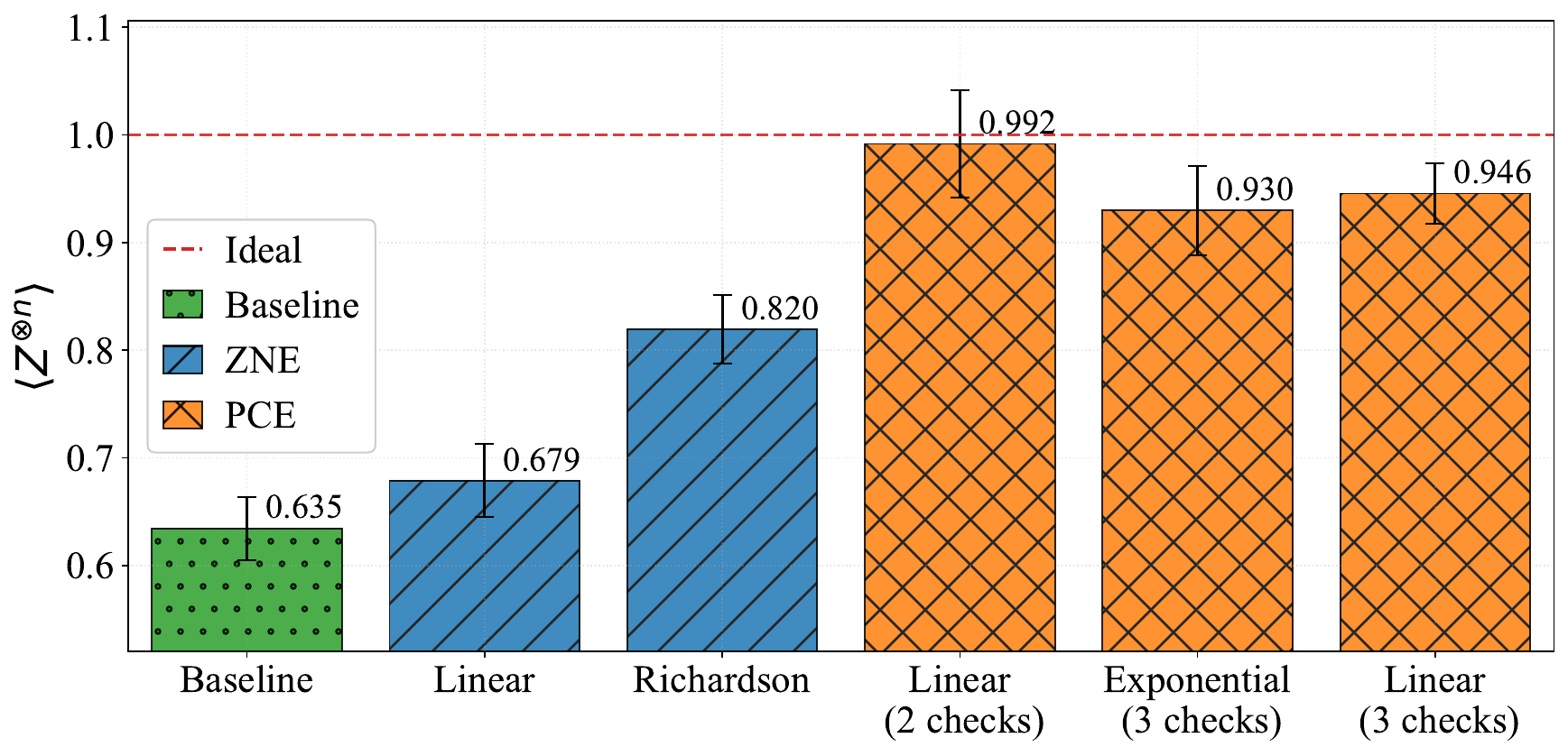}
\caption{Quantum hardware. Expectation values for 4-qubit mirrored random Clifford circuits with total circuit depth of $50$. Comparing baseline, ZNE-linear, ZNE-Richardson, and PCE methods (2 and 3 checks) with exponential and linear fits. Error bars indicate $\pm 1$ SD. The numbers on top of each bar is the corresponding average expectation value over $5$ random circuits.}
\label{fig:hw_results_d=25}
\end{figure}

The results are shown in Figure~\ref{fig:hw_results_d=25} for pre-mirroring circuit depths of $25$ (total depth of $50$ after mirroring). We see that each PCE implementation provides significant improvement over the baseline, ranging from $46.5\%$ to $56.2\%$ improvement, with an accuracy of $99.2\%$. In comparison, ZNE achieves an accuracy of up to $82\%$, which is a $29.1\%$ improvement. 

The error bars show that extrapolation from $2$ checks exhibits greater variability than from $3$ checks, as expected, yet achieves slightly better average performance. This suggests that the device’s error rates may have been too high for reliable estimation with $3$ checks, since the reduced post-selection rate makes accurate sampling more difficult. The issue could be mitigated by using a lower-noise device or by increasing the number of samples.

\subsection{Application to Classical Shadow Tomography (classical simulation of PCE vs RS Estimation)}
\label{sec:Results}

% \subsection{Experimental Setup}

\begin{figure}[ht]
\centering
\includegraphics[width=0.48\textwidth]{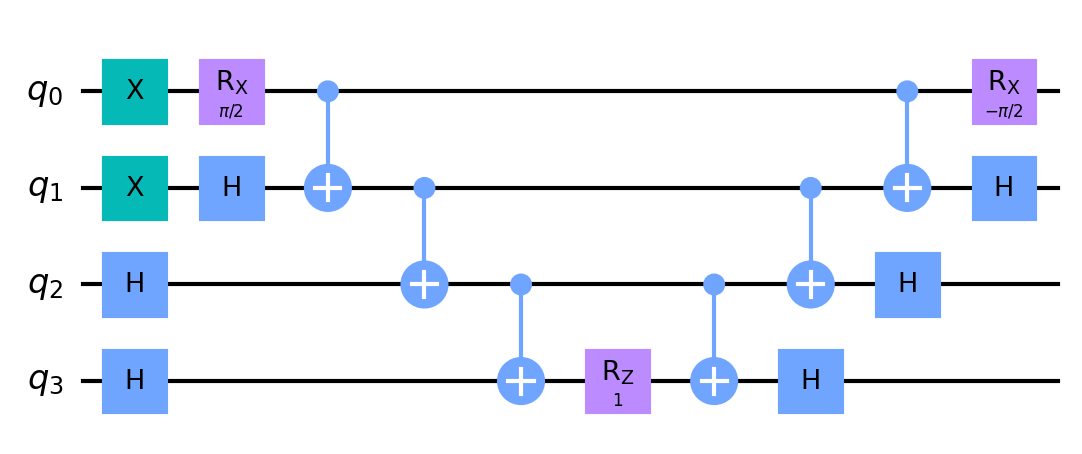}
\caption{State preparation circuit for the H$_2$ molecule using $4$ qubits. The variational parameters were optimized from a prior noiseless VQE calculation.}
\label{fig:hydrogen_trial_circuit}
\end{figure}

\textbf{Experimental setup.} For the following set of classical simulation experiments, we applied PCE towards error mitigating the expectation value of tensor products of Pauli operators derived from classical shadow tomography for VQE circuits. We tested on two VQE circuits: a $4$-qubit circuit for modeling the H$_2$ molecule and an $8$-qubit circuit for H$_2$O. The H$_2$ circuit architecture is shown in Fig.~\ref{fig:hydrogen_trial_circuit}. In the classical shadow protocol applied in these experiments, each VQE state preparation circuit is followed by a random unitary from the global Clifford group $\text{Cl}(2^n)$ and subsequent $Z$-basis measurements.

% \begin{figure}[ht]
% \centering
% \includegraphics[width=0.4\textwidth]{global_clifford.png}
% \caption{Example of a random global Clifford circuit appended to the hydrogen VQE circuit for classical shadow tomography.}
% \label{fig:global_clifford_circuit}
% \end{figure}

A widely used type of ansatz for solving the eigenvalue problem of molecular and solid-state systems is the unitary coupled-cluster (UCC)~\cite{romero2018strategies} circuit, which is inspired from coupled-cluster theory~\cite{bartlett2007coupled}. UCC ansatz is systematically improvable but suffers from long circuit depth, which makes it hard for near-term quantum hardware. A workaround is the qubit coupled-cluster (QCC)~\cite{Ryabinkin2018} ansatz, which reduces the circuit depth significantly by a pre-screening procedure. For the circuits used in this work, we adopted the QCC circuits with variational parameters optimized from a prior noiseless VQE calculation.

The VQE circuits were evaluated under four conditions: ideal noiseless execution, noisy execution without mitigation, noisy execution with RS estimation, and noisy execution with PCE using varying numbers of check layers.

For a given state preparation circuit, we generated $10,000$ shadow circuits by appending random global Clifford operations to the VQE circuit. We then collected $100$ computational basis measurement samples from each circuit. We post-processed the measurement data by estimating observables across varying numbers of shadow circuits $N$: $100$, $400$, $1000$, $4000$, and $10000$. For each value of $N$, we sampled $N$ shadow circuits from the total pool and partitioned them into 20 equally sized sets, each containing $N/20$ circuits. To reduce the effect of outliers, we computed the mean observable estimate for each set and then took the median of these means. This process followed the procedure outlined in Algorithm 1 from \cite{Huang_2020}.

In this implementation of PCS within classical shadow tomography, the measurement is conducted exclusively in the $Z$-basis. Therefore, errors on the phase do not impact the output, limiting the utility of checks beyond the number of qubits in the compute circuit. Accordingly, we used the first $3$ check expectation values to extrapolate to a maximum of $4$ checks for the H$_2$ circuit and used the first $4$ checks to extrapolate to $8$ checks for the H$_2$O circuit.

% \subsection{Fully connected topology}

\textbf{Fully connected topology.} We begin by examining a fully connected topology where each gate, including single-qubit and two-qubit gates, is subjected to depolarizing noise. The error rates are denoted as $p_1$ for single-qubit gates and $p_2$ for two-qubit gates. We allow the depolarizing noise to affect all gates in the circuit, including the checks $L_n$ and $R_n$.

\begin{figure}[!htb]
    % \centering
    \includegraphics[width=0.48\textwidth]{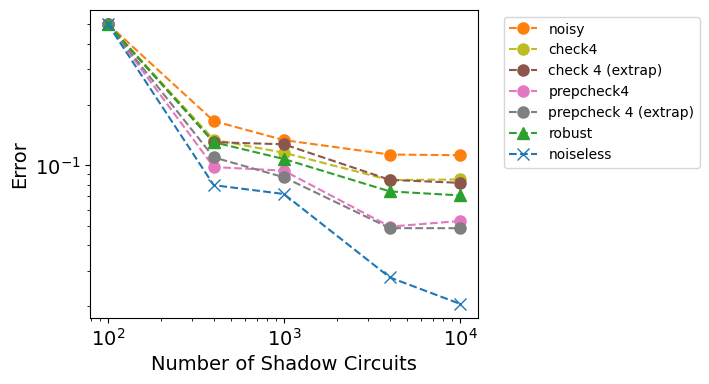}
    \caption{
Classical simulation. Error rates for a 4-qubit H$_2$ circuit with full connectivity under a depolarizing error channel with \( p_1 = 0.002 \) and \( p_2 = 0.02 \). Results compare protecting only the Clifford portion (check) versus the entire circuit (prepcheck). `Check 4 (extrap)' and `prepcheck 4 (extrap)' are the extrapolated expectation values from the first three check implementations.}

    \label{fig:4-qubit_full_connect}
\end{figure}

% Discussion of 4-qubit results
As mentioned in Section~\ref{Sec:Background}, in the context of a Clifford circuit, it is always possible to identify a Pauli check $L_n$ corresponding to any given $R_n$ such that the condition in Eq.~(\ref{eq:PCS}) is satisfied. Figure~\ref{fig:4-qubit_full_connect} shows the results for the scenario where only the global Clifford portion of the classical shadow protocol is protected, labeled `Check 4' and `Check 4 (extrap)' in the legend. A distinct advantage of PCS over RS estimation (labeled `robust') is its capability to extend protection to the state preparation portion of the circuit, in addition to the global Clifford. This extended protection resulted in PCE outperforming RS estimation when the entire circuit was protected (Figure~\ref{fig:4-qubit_full_connect}). The extrapolated check provided similar performance to its implemented counterpart, suggesting that the extrapolation technique accurately predicted the expectation values for the maximum check configuration. Similar accuracy was observed across all tested scenarios, including the IBM mock backends.

% Clifford and state prep + Clifford protection for 4-qubit hydrogen circuit, depolarizing error

\begin{figure}[!htb]
    \centering
    \begin{subfigure}[b]{0.48\textwidth}
        \captionsetup{justification=raggedright,singlelinecheck=false,font=small,labelfont=bf,labelsep=quad}
        \caption{}
        \includegraphics[width=\textwidth]{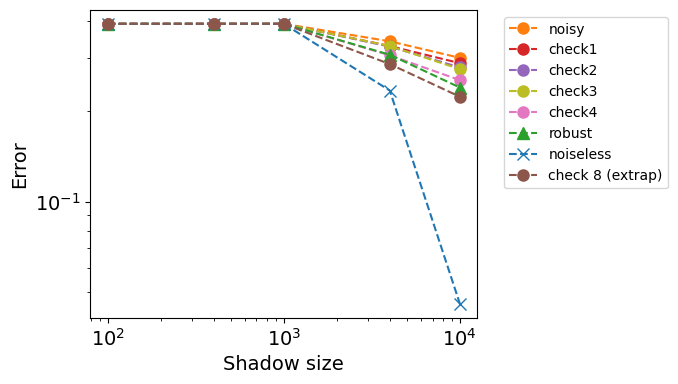}
        \label{fig:H2O_even_noise}
    \end{subfigure}
    \hfill
    \begin{subfigure}[b]{0.48\textwidth}
        \captionsetup{justification=raggedright,singlelinecheck=false,font=small,labelfont=bf,labelsep=quad}
        \caption{}
        \includegraphics[width=\textwidth]{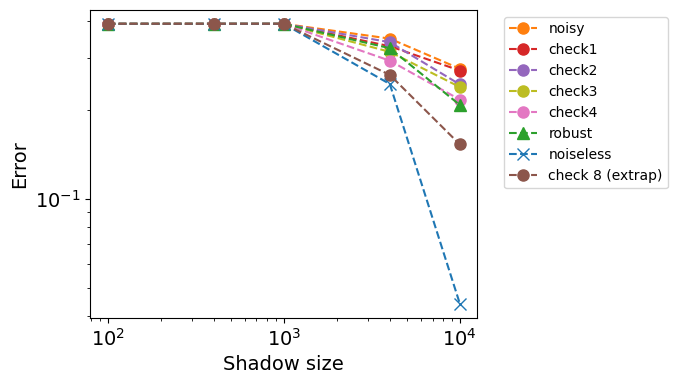}
        \label{fig:H20_uneven_noise}
    \end{subfigure}
    \caption{Classical simulation. Fully connected 8-qubit H$_2$O circuit with homogeneous and heterogeneous noise distributions. (a) Homogeneous depolarizing error with $p_1 = 0.002$ and $p_2 = 0.02$ across all qubits. (b) Heterogeneous depolarizing error with $p_2 = 0.02 \pm 0.005$ and $p_1 = 0.002 \pm 0.0005$.}
    \label{fig:8-qubit_H2O_compound}
\end{figure}

For the 8-qubit H$_2$O circuit, PCE extrapolated to eight check layers achieved superior fidelities compared to RS estimation, as shown in Figure~\ref{fig:8-qubit_H2O_compound}. This performance advantage becomes particularly relevant when considering the noise assumptions underlying each method. RS estimation assumes uniform noise distribution across qubits (Assumption A1 in \cite{Chen_2021}), an assumption that rarely holds in real hardware where error rates vary significantly between qubits, even for identical gate operations \cite{Aseguinolaza_2024}. In contrast, PCE makes no assumptions about noise uniformity, allowing it to naturally accommodate the heterogeneous error distributions found in physical devices.

To test this theoretical advantage experimentally, we simulated inhomogeneous noise by applying a Gaussian distribution of error rates across qubits, with mean values of $p_1 = 0.002$ and $p_2 = 0.02$ and standard deviations of $0.0005$ and $0.005$, respectively. As shown in Figure~\ref{fig:H20_uneven_noise}, PCE's performance gap over RS estimation widened substantially under these heterogeneous conditions compared to the homogeneous case (Figure~\ref{fig:H2O_even_noise}). This enhanced performance under realistic noise conditions suggests that PCE may be better suited for practical quantum computing environments where noise heterogeneity is the norm.

% highlight that figure 5 shows that extrap is close to real check 4

% \subsection{IBM device topology}
\label{sec:IBM device}

\textbf{IBM device topology.} This subsection analyzes PCE performance using realistic noise models and circuit topologies from IBM quantum devices, specifically mock backends provided by Qiskit. The analysis focuses on the case where PCE protects the entire 4-qubit H$_2$ shadow circuit (both state preparation and Clifford portions) and compares the results to RS estimation.

\begin{figure}[!htb]
    \centering
    \begin{subfigure}[b]{0.48\textwidth}
        \captionsetup{justification=raggedright,singlelinecheck=false,font=small,labelfont=bf,labelsep=quad}
        \caption{}
        \includegraphics[width=\textwidth]{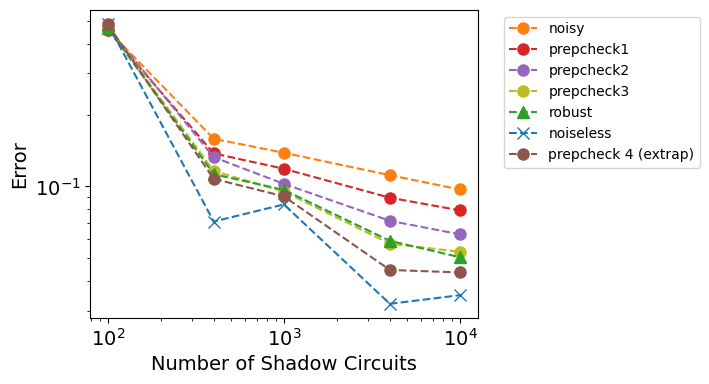}
        \label{fig:cairo_stateprep}
    \end{subfigure}
    \hfill
    \begin{subfigure}[b]{0.48\textwidth}
        \captionsetup{justification=raggedright,singlelinecheck=false,font=small,labelfont=bf,labelsep=quad}
        \caption{}
        \includegraphics[width=\textwidth]{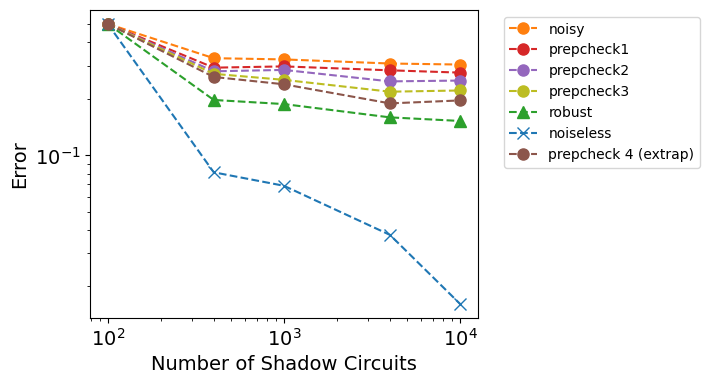}
        \label{fig:melbourne_stateprep}
    \end{subfigure}
    \caption{Classical simulation. 4-qubit H$_2$ circuits under IBM device-specific constraints. Results shown for (a) Cairo backend and (b) Melbourne backend, with `prepcheck' denoting protection of both state preparation and Clifford portions of the circuit.}
    \label{fig:4-qubit_hydrogen_ibm_devices}
\end{figure}

Figure~\ref{fig:4-qubit_hydrogen_ibm_devices} presents PCE performance on two IBM mock backends: Cairo and Melbourne. The backend-specific error rates were applied to the entire circuit, including the controlled Pauli checks. When protecting both the state preparation and Clifford portions, PCE extrapolated to four layers achieved performance comparable to RS estimation. The error reduction curves maintained consistent trends across all check layers, including the extrapolated fourth check, indicating that the extrapolation model remains valid for non-depolarizing noise models characteristic of real hardware.

Testing across multiple IBM mock backends revealed a consistent pattern: PCE performed comparably to RS estimation, occasionally achieving slightly improved fidelities when the entire circuit was protected. The performance differential between PCE and RS appears to depend primarily on circuit topology rather than specific noise characteristics. This dependence on the topology is evident when comparing the fully connected results (Figure~\ref{fig:4-qubit_full_connect}) with the device-constrained topologies (Figure~\ref{fig:4-qubit_hydrogen_ibm_devices}). Under full connectivity, PCE with $4$ checks sandwiching the entire circuit substantially outperformed RS estimation, whereas on constrained IBM topologies, the two methods yielded similar performance. Note that ancilla placement was not optimized for these experiments, likely degrading PCE performance through unnecessary two-qubit gate overhead. Although connectivity constraints can lead to greater variability in performance, applying Pauli-X checks to ancilla qubits and post-selecting noise-free ancilla results helps mitigate this issue \cite{single-shot_error_mitigation}, showing significant improvements over unprotected ancilla qubits \cite{Gonzales_2025}. This remains for future work. 

% Clifford+state prep IBM device (4-qubit)

\section{Conclusions and Future Directions}

This paper introduces a novel error mitigation technique that extends upon PCS by extrapolating the Pauli checks. This approach helps achieve similar fidelities associated with the maximum number of checks while avoiding practical issues such as additional noise and the exponentially decreasing post-selection rate that arise from physically adding more checks. In our experiments, PCE outperformed ZNE on larger random Clifford circuits, using a consistent exponential extrapolation model across all circuits while ZNE required different scaling factors and extrapolation methods for optimal performance on each circuit. We were also able to show improved performance on real hardware, demonstrating its practical applicability on the IBM Kingston backend. We additionally applied PCE to mitigate errors in classical shadows and demonstrated that PCE can offer equal or superior performance compared to the current state-of-the-art classical shadow error mitigation scheme. Importantly, PCE achieves this performance without requiring a calibration step, significantly reducing the overall number of samples needed for implementation compared to RS estimation.

The primary trade-off to consider when using PCE is its restriction to Clifford or near-Clifford circuits where valid Pauli checks exist. Despite this, PCE can still protect Clifford subcircuits embedded in non-Clifford circuits. From our experiments, we can see the utility of this approach towards mitigating classical shadow circuits. This partial protection approach could be extended to other quantum algorithms with identifiable Clifford blocks. For instance, many variational algorithms contain Clifford entangling layers between parameterized rotation layers, and quantum phase estimation circuits include substantial Clifford components in their controlled-unitary structure. Future research could investigate optimal strategies for identifying and protecting maximal Clifford subcircuits, potentially combining PCE on Clifford portions with other error mitigation techniques for non-Clifford gates. 

% Optimization of extapolation model
This paper utilized both linear and exponential ansatzes to extrapolate expectation values. The exponential ansatz, $E(m) = ab^m + c$, where $a$, $b$, and $c$ are fitting parameters, was derived in Section~\ref{sec:exp_ansatz} using the Markovian model described in Ref.~\cite{Vandenberg_2023SingleShotErrMitigByCohPauliChecks}. In simulated experiments with depolarizing noise, the exponential model significantly outperformed the linear model. However, on quantum hardware, the exponential model produced results comparable to the linear model, with the parameter $b$ frequently converging near 1, effectively yielding a linear fit. The hardware experiments used only a few data points for fitting, limiting the exponential model's potential advantages. With larger circuits and more implemented checks, the exponential model may demonstrate superior performance, like was shown in the simulation results. 

For the experiments in this paper, the optimization bounds for the parameter $b$ in Eq~\ref{eq:exp_ansatz} were set to $0.6 \leq b \leq 1.2$, while parameters $a$ and $c$ remained unbounded. Optimal parameter constraints likely depend on multiple factors: the number of implemented checks, circuit structure, noise characteristics, etc. Future work could explore adaptive parameter selection and alternative extrapolation models, including higher-order polynomial ansatzes, to further improve PCE performance.

Future research could explore several directions to improve PCE performance. Limited device connectivity can degrade PCS performance because non-adjacent ancilla and target qubits require additional noisy gates. Recent work by Martiel and Javadi-Abhari~\cite{martiel2025lowoverheaderrordetectionspacetime} addresses this challenge on superconducting devices through an efficient algorithm for finding checks that respect connectivity constraints while maximizing circuit coverage. This algorithm could significantly improve PCE performance on connectivity-limited devices.
Alternatively, future work could explore PCE on quantum devices with full connectivity, such as neutral atom and trapped-ion systems, which would eliminate connectivity-induced overhead entirely. Additional techniques like recursive checks (checks on ancilla qubits) \cite{Gonzales_2025} or single-sided checks~\cite{single-shot_error_mitigation} could also improve performance across all device types.

% Additionally, to help reduce noise in the checks and imporve the reliability, we could implement the checks

% =========================
% Data & Code Availability
% =========================
% \section*{Data Availability}
% \href{https://www.nature.com/npjqi/for-authors-and-referees/submission-guidelines#format-manuscripts}{manuscript guidelines}

% \section*{Code Availability}

\section{Acknowledgements}
QL, JL, BH, ZHS, and AG acknowledge support by the Q-NEXT Center. NH acknowledge support from NSF award CCF-2119069 and Northwestern University's McCormick School of Engineering. AG and ZHS also acknowledge support by the U.S. Department of Energy (DOE) under Contract No. DE-AC02-06CH11357, through the Office of Science, Office of
Advanced Scientific Computing Research (ASCR) Exploratory Research for Extreme-Scale Science and Accelerated Research in Quantum Computing. We gratefully acknowledge the computing resources provided on Bebop, a high-performance computing cluster operated by the Laboratory Computing Resource Center at Argonne National Laboratory. This research used resources of the Oak Ridge Leadership Computing Facility, which is a DOE Office of Science User Facility supported under Contract DE-AC05-00OR22725. We acknowledge the use of IBM Quantum services for this work through an allocation awarded through the IBM Quantum Credits Program. The views expressed are those of the authors, and do not reflect the official policy or position of IBM or the IBM Quantum team.

\bibliographystyle{apsrev4-1}
\bibliography{refs}

\vfill

\small

\framebox{\parbox{\linewidth}{
The submitted manuscript has been created by UChicago Argonne, LLC, Operator of 
Argonne National Laboratory (``Argonne''). Argonne, a U.S.\ Department of 
Energy Office of Science laboratory, is operated under Contract No.\ 
DE-AC02-06CH11357. 
The U.S.\ Government retains for itself, and others acting on its behalf, a 
paid-up nonexclusive, irrevocable worldwide license in said article to 
reproduce, prepare derivative works, distribute copies to the public, and 
perform publicly and display publicly, by or on behalf of the Government.  The 
Department of Energy will provide public access to these results of federally 
sponsored research in accordance with the DOE Public Access Plan. 
http://energy.gov/downloads/doe-public-access-plan.}}

\end{document}